\documentclass{article}
\usepackage[T1]{fontenc} % add special characters (e.g., umlaute)
\usepackage[utf8]{inputenc} % set utf-8 as default input encoding
\usepackage{ismir,amsmath,cite,url}
\usepackage{graphicx}
\usepackage{color}

\usepackage[bookmarks=false,hidelinks]{hyperref}
\usepackage{caption}
\usepackage{subcaption}

\newcommand\para[1]{{\vspace{1mm}\noindent\textbf{#1}.}}

\definecolor{green}{rgb}{0, 0.8, 0}
\definecolor{red}{rgb}{0.8, 0, 0}
\title{Visual Overviews for Sheet Music Structure}

\multauthor
{Frank Heyen \hspace{1cm} Quynh Quang Ngo \hspace{1cm} Michael Sedlmair} { \bfseries{}\\
VISUS, University of Stuttgart, Germany\\
{\tt\small \{frank.heyen, quynh.ngo, michael.sedlmair\}@visus.uni-stuttgart.de}
}

\def\authorname{F. Heyen, Q. Q. Ngo, and M. Sedlmair}

\begin{document}

\maketitle
\begin{abstract}
% The abstract should be placed at the top left column and should contain about 150-200 words.

We propose different methods for alternative representation and visual augmentation of sheet music that help users gain an overview of general structure, repeating patterns, and the similarity of segments.
To this end, we explored mapping the overall similarity between sections or bars to colors.
For these mappings, we use dimensionality reduction or clustering to assign similar segments to similar colors and vice versa.
To provide a better overview, we further designed simplified music notation representations, including hierarchical and compressed encodings.
These overviews allow users to display whole pieces more compactly on a single screen without clutter and to find and navigate to distant segments more quickly.
Our preliminary evaluation with guitarists and tablature shows that our design supports users in tasks such as analyzing structure, finding repetitions, and determining the similarity of specific segments to others.
\end{abstract}
\section{Introduction}\label{sec:introduction}

\begin{figure*}
 \centerline{
 \includegraphics[width=\linewidth]{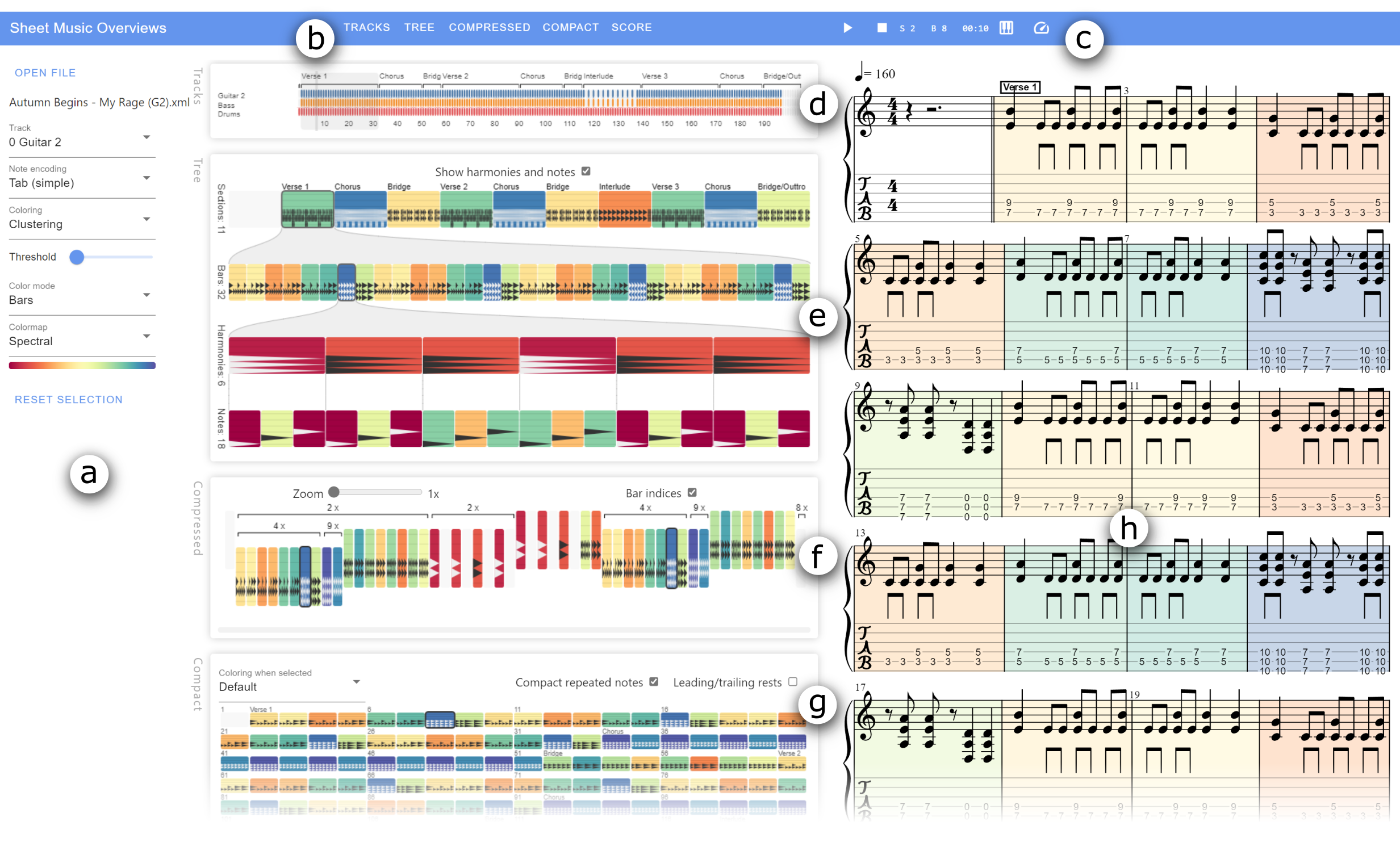}}
 \caption{
     Screenshot of our design with all views (cut):
    a) data and visualization options,
    b) view selection,
    c) player,
    d) instrument/track overview,
    e) structure hierarchy,
    f) compressed repetitions,
    g) compact sheet, and
    h) complete score.
 }
 \label{fig:teaser}
\end{figure*}

Common music notation can be considered as a special visual encoding to convey music, including instructions on how to perform it.
Despite its compactness and detailed information, a music sheet is hard to analyze for novice musicians~\cite{fuerst2020augmenting}.
Moreover, it contains lots of information that is hard to display at once without visual clutter or getting too small -- getting an overview is tricky.
When pieces contain repeating sections such as a chorus, much information is redundant.
Even with abbreviations that denote repetitions, such as a double bar with colon, da capo, or al segno, a complex structure can lead to tedious navigation.
%

%Other research
Recent work~\cite{miller2019augmenting, miller2022augmenting, fuerst2020augmenting} strove to enrich notation to better convey music-theoretical information and patterns.
In this paper, we instead focus on  quickly gaining an overview of structures such as similarities and repetitions.
This overview is meant to support learning, or teaching a music piece, as musicians often have to remember which segment of a piece they have to play when and how often, information that can be obscured in classical sheet music notation.
According to the visualization principle ``eyes beat memory''~\cite{munzner2014book_visualization}, we argue that the current notation leaves room for further optimization.

Toward this goal, we  propose to visually enrich sheet music by mapping calculated similarities among segments of the sheet music, such as sections or bars (measures), to colors.
To fit the notes of a whole piece onto a screen while remaining legible, we propose compact alternative encodings to common music notation that allow for overview and easier navigation without having to scroll or change pages.
This work focuses on guitar tablature of Western music, which is easier to represent compactly and often features more repeating parts than other kinds of sheet music.
However, we argue that our general method of mapping similarities to color can also help with other kinds of music.
We conducted a preliminary qualitative evaluation through pair analytics with four guitarists.
The results indicate that our design supports tasks such as summarizing structure, finding repetitions, and analyzing similarity.

%% summary contribution
In summary, we contribute
1) the exploration of novel representations of sheet music for easier overviews, specifically a method for mapping similarity among components of a music sheet to color, and
2) a preliminary pair analytics study with four guitarists.
We further provide source code and a web app where users can try their own sheet music in MusicXML~\cite{good2001musicxml} at \href{https://visvar.github.io/sheetmusic-overviews}{visvar.github.io/sheetmusic-overviews}.

% \vspace{-0.5mm}

%%------------------------------------------------------------------------
\section{Related Work}

Similarity in music concerns many dimensions such as cognition, perception, tempo, pitch, and more.
Therefore, existing metrics use different approaches, including a continuous representation of notes\cite{Valle2015SymbolicMS},
a geometrical metric~\cite{deHaas2013},
shapes of curves~\cite{Urbano2010:using}, and
a graph-based metric for harmony~\cite{simonetta2018sym_mus_sim_graphbased}.
Janssen et al.~\cite{Janssen2015ACO} evaluated melodic similarity metrics using human annotators and a survey~\cite{SymbolicMelodicSimilaritySurvey2016} defined eight criteria for symbolic melodic similarity.
The overall aim of the above work is to query pieces in a database.
In contrast, our work focuses on supporting the structural analysis of a single piece, by visualizing similarities within it.
While our design allows integrating any metric, its main purpose is demonstrating how visualization can support sheet music analysis.
We thus use a simpler symbolic metric to instantiate our design.

There is a broad range of music-related visualization~\cite{khulusi2020survey, LimaSurvey2021}, including structure~\cite{watanabe2003brass, wattenberg2002arcdiagrams, li2016music, cantareira2016moshviz, savelsberg2021visualizing, muller2012scape, hayashi2011colorscore, bergstrom2007isochords, snydal2005improviz}.
However, some visual encodings, such as one based on Tonnetz~\cite{bergstrom2007isochords}, require knowledge of music theory.
Similar to our approach, \textit{MoshViz}~\cite{cantareira2016moshviz} focuses on visual analysis in an overview-detail fashion, but without considering perceptual or cognitive aspects.
Our design allows encoding any similarity metric (mathematical or perceptual) with colors, to make sheet music easier to understand.
Closest to our work is a structure visualization that uses dimensionality reduction (DR) to map audio features to color~\cite{savelsberg2021visualizing}, which inspired us to design a similar mapping for sheet music.

% \para{Augmented Sheet Music}
%
Augmented sheet music adds visual components to common music notation to increase expressiveness.
Related work augments a music piece with radial note histograms, to facilitate analyzing harmonic patterns~\cite{miller2019augmenting}, or visualizes rhythm through color-coded sunbursts~\cite{fuerst2020augmenting}.
Miller et al.~\cite{ miller2022augmenting} combined both approaches, but do not address supporting performance preparation tasks.
Only little research supports instrument learning and composition.
Bunks et al.~\cite{bunks2022jazz} use color for reference keys on a tabular layout to support jazz improvisation.
Others augment sheet music with lines and ellipses to support error detection in composition~\cite{deprisco2017understanding} or pianists in identifying mistakes while practicing~\cite{asahi2018toward:piano:support, hori2019piano:hmm}.
In this work, we also support learning by aiding music reading before and during practice.
%

%%------------------------------------------------------------------------
\section{Design}

We first introduce the tasks we want to support with our approach.
Then, we describe how we compute similarities and map them to colors and how we represent sheet music visually (\autoref{fig:teaser}).

\subsection{User Tasks}

\newcommand{\task}[1]{$\mathbf{T_{#1}}$}

Our overall goal is to improve the efficiency of reading sheet music, by sparing users the need to search for certain segments or memorize patterns.
We want to reveal potentially interesting patterns that are hard to infer from the bare sheet music itself but could be helpful for better understanding or practicing a piece.
More specifically, we want to support the following tasks:
(\task{1}) understand the \textit{overall structure} of a piece,
(\task{2}) detect repetitions, which means to spot \textit{where} something repeats \textit{how often}, and detect \textit{repetitions nested within} repetitions,
(\task{3}) \textit{compare} multiple segments regarding their similarity.

\subsection{Color Mappings}

\para{Similarity metrics}
Our approach works with any metric that takes notes and returns a scalar similarity score.
We compare non-overlapping segments of the piece, which can be bars, pre-defined sections read from the MusicXML file, or the result of an automatic segmentation (the latter is not implemented).

In our related work section, we discussed existing similarity metrics for symbolic music.
Some of these metrics do not support polyphony or require complete scores or additional annotations (such as chords) or assumptions on musical meaning.
Metrics that are based on western tonal harmony~\cite{dehaas2009modeling, dehaas2011harmtrace} would also not generalize to various cultures and genres.
%
% what we did
Therefore, we designed the following simple but robust algorithm:
First, the notes of a segment are sorted by their start time and those with equal start time by pitch ascending.
Mapping each note to its pitch then results in one sequence of integers for each segment.
We then compute a similarity matrix by calculating the \textit{Levenshtein distance}~\cite{levenshtein1966binary} for all possible pairs of segments, equals the minimum number of pitches one would have to insert, delete, or replace, to transform the first sequence into the second.

We further compute similarities between all sets of notes that have the same start time, which we refer to as harmonies.
For these sets, we only use the notes' pitch class (disregarding octaves) and compute the \textit{Jaccard index}~\cite{jaccard1901distribution}, which equals the ratio of intersection over union.

\begin{figure}[htp]
  \centering
  \mbox{} \hfill
  \includegraphics[width=\linewidth]{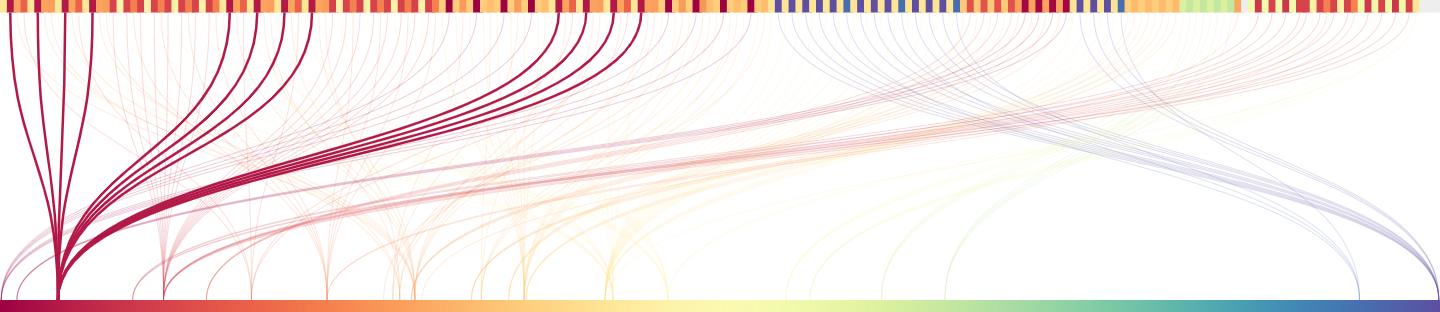}
  \hfill \mbox{}
  \caption{\label{fig:mapping}%
    Example of similarity-based color mappings.
    \textit{Top:} rectangles represent all bars of a piece, from left to right in the order they occur.
    \textit{Bottom:} a color scale.
    Curves connect each bar to its color.
    In this figure, the curves of a single color are highlighted through a stronger opacity to show how they connect to identical bars.
   }
\end{figure}

\begin{figure}[htp]
  \centering
  \mbox{} \hfill
  \includegraphics[width=\linewidth]{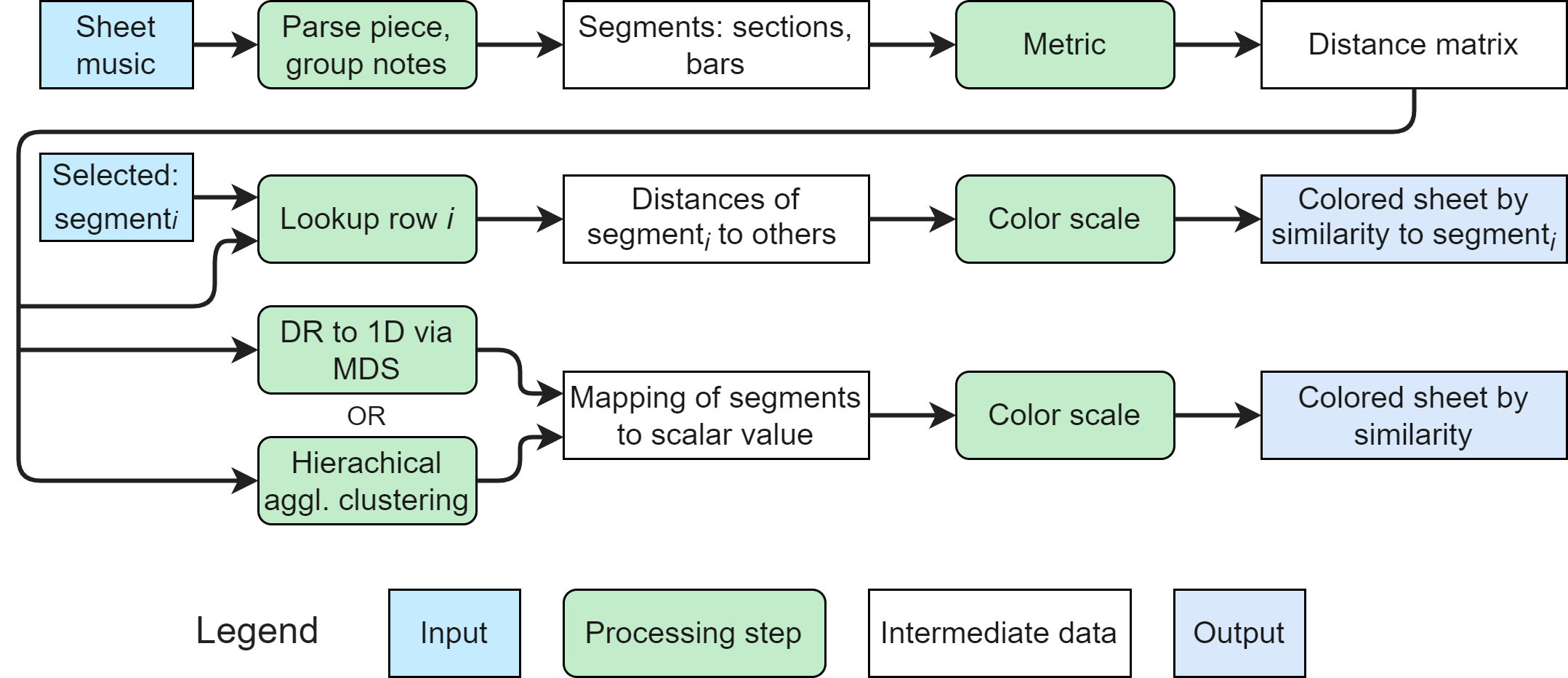}
  \hfill \mbox{}
  \caption{\label{fig:pipeline}%
    We compute similarity-based colors for extracted segments via direct comparison, dimensionality reduction, or clustering.
    A \textit{segment} can be any sequence of notes in the piece, such as a bar or a pre-defined section.
    The \textit{selected segment} is chosen by the user to compare it to all others.
   }
\end{figure}

\para{Mapping}
Once we have a similarity matrix, we can create a color mapping that respects these similarities.
We explored three alternative methods that use either a one-to-many comparison, dimensionality reduction, or hierarchical clustering.
\autoref{fig:mapping} shows an example for our similarity-based color mapping, \autoref{fig:pipeline} summarizes our different mapping strategies.

The first method colors bars by their similarity to a selected bar.
This selection is made by the user or automatically when playing the piece, where the currently played bar is selected.
To obtain colors, we linearly map the similarities to a color scale.
Another mode only colors bars that the metric considers identical to the selected one, allowing users to quickly spot where and how often it repeats.

Our second method uses dimensionality reduction (DR), a method commonly used to transform data from a high-dimensional space to a lower-dimensional one.
Usually, the target space is two- or three-dimensional, such that data points can be shown on a screen.
We instead project onto a one-dimensional space that we can then linearly map to a color scale.
As we do not have concrete positions in a space, but only the similarities between them, we chose \textit{multi-dimensional scaling} (MDS)~\cite{kruskal1964multidimensional} that accepts a similarity matrix as input.
Furthermore, MDS optimizes the computed projection to preserve these similarities, leading to a coloring optimized for these.

As an alternative to MDS, we designed a method that clusters similar segments together and then assigns each cluster one color such that similar clusters have similar colors.
Using our similarity matrix, we compute \textit{hierarchical agglomerative clustering}, which gives us a binary tree.
We then sort the tree's leaves from left to right, as leaves that are closer together are more similar, and map them in this order to a color scale.
Compared to the above method using MDS, the resulting colors are easier to distinguish but represent similarities less accurately.
Using a similarity threshold, users can steer the number of clusters and therefore colors, to choose a trade-off between detail and overview.

\begin{figure}[htp]
  \centering
  \mbox{} \hfill
  \includegraphics[width=0.95\linewidth]{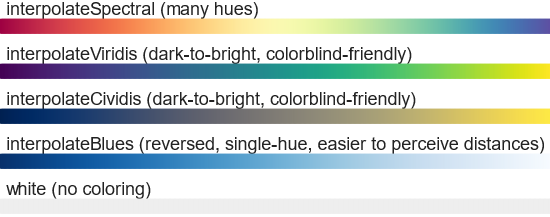}
  \hfill \mbox{}
  \caption{\label{fig:color_scales}%
    Some of the included color scales, taken from \href{https://github.com/d3/d3-scale-chromatic}{D$^3$}\cite{bostock2011d3}.
    The choice depends on the current task and individual limitations of the user's color vision.
   }
\end{figure}

\para{Color scales}
% colormaps may depend on task
Research on perception proposed a range of color scales specifically designed for visualization.
Since there are different irreconcilable goals, no scale is appropriate for all tasks.
While multi-hue scales such as rainbows have been criticized~\cite{borland2007rainbowstillharmful}, they have been shown to work well for some circumstances~\cite{reda2021rainbows, reda2021color, wattenberg2007chromograms}.
% cividis: https://journals.plos.org/plosone/article?id=10.1371/journal.pone.0199239
For users with a color vision deficiency, scales with fewer hues and, therefore, less discernible colors can be used, such as cividis~\cite{nunez2018cividis}.
When color is used to compare different values or intervals, a color scale needs to accurately represent similarities between values.
For this task, single-hue scales or interpolations between two hues are appropriate but further reduce the number of discernible colors.
Although the number of distinguishable colors is small, there are enough for our use case, as the number of different segments in a piece is limited.
Since colors are distributed by similarity, indistinguishable colors should only be assigned to very similar segments.
To accommodate different user needs, we choose a broad multi-hue scale (spectral) as default but also provide more accessible ones; for direct comparison, we choose a single hue scale (blues) (\autoref{fig:color_scales}).

%%-----------------------------------------------------------------------
\subsection{Visual Encodings}

\begin{figure}[tp]
  \centering
  \mbox{} \hfill
  \includegraphics[width=\linewidth]{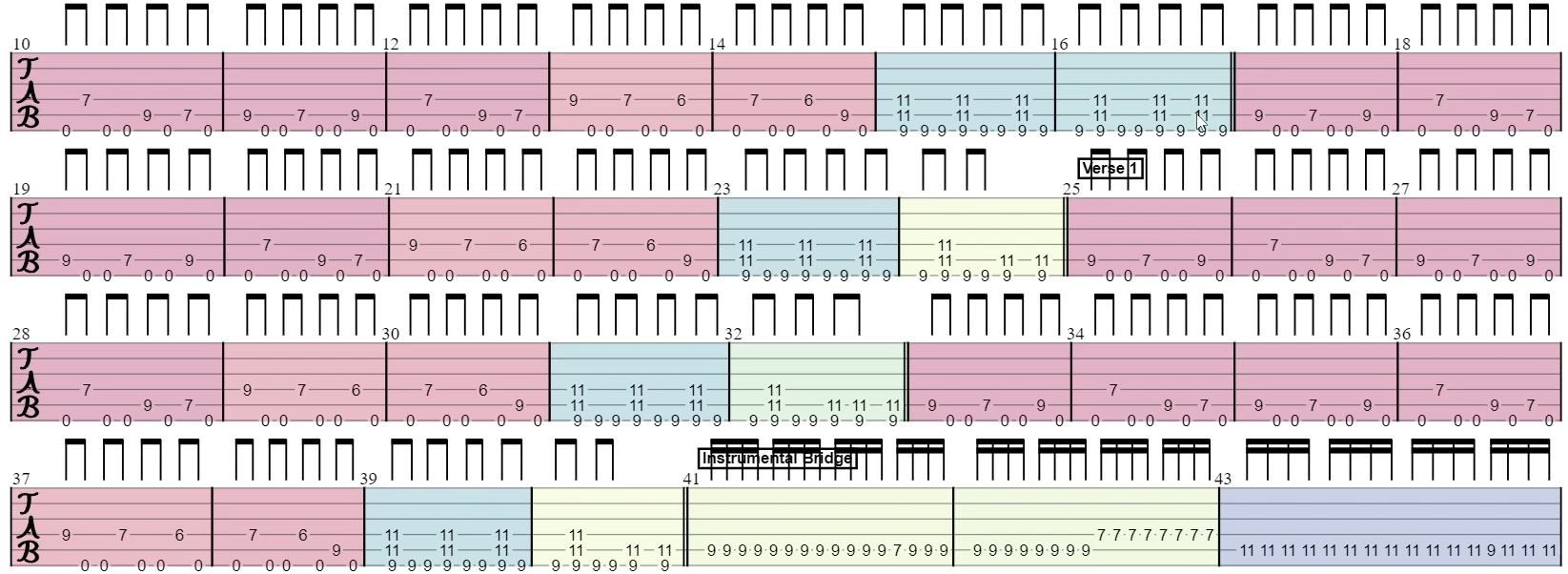}
  \hfill \mbox{}
  \caption{\label{fig:differentendings}%
   A pattern of repetitions with different endings.
   Bar 16 is colored blue, 24 and 40 are the same yellow, and bar 32 is green.
   }
\end{figure}

\para{Layout enrichment for common music notation}
We designed several visual encodings to address different tasks and reveal different kinds of patterns.
The most straightforward encoding is to display the full sheet music as the common notation that is familiar to musicians and represents the complete information.
We enrich this display by adding colored, semi-transparent rectangles on top of the segments, for example, one for each bar (\autoref{fig:differentendings}).
The reduced opacity makes colors brighter than in other views but improves the readability of the notation.
This coloring helps more quickly see where a bar repeats (\task{2}), as the user only has to compare bars with similar colors (\task{3}).
Even when two different bars were assigned similar colors, this process allows for ruling out many others.
This encoding suffers from the same limitations as non-colored sheet music.
Due to its highly detailed nature, fitting the complete piece on the screen at once would lead to small and illegible visuals.
Therefore, we designed simplified, filtered, and compressed alternatives, which we explain in the following paragraphs.

\begin{figure}[htp]
    \centering
      \begin{subfigure}[b]{0.32\linewidth}
         \centering
         \includegraphics[width=\linewidth]{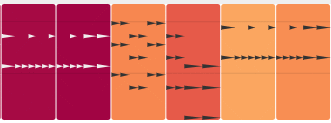}
         \caption{Piano roll.}
         \label{fig:note_encoding_pianoroll}
     \end{subfigure}
      \begin{subfigure}[b]{0.32\linewidth}
         \centering
         \includegraphics[width=\linewidth]{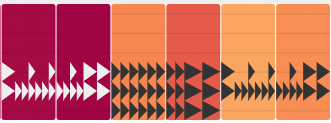}
         \caption{Tab (simple).}
         \label{fig:note_encoding_tabsimple}
     \end{subfigure}
      \begin{subfigure}[b]{0.32\linewidth}
         \centering
         \includegraphics[width=\linewidth]{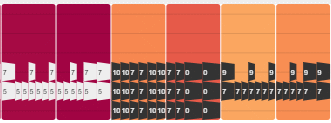}
         \caption{Tab with frets.}
         \label{fig:note_encoding_tabfrets}
     \end{subfigure}
    \caption{
    Note display:
    a) Piano rolls can represent any music but lack additional information.
    b, c) Tablature either simplified or with fret numbers.
    Notes are drawn in black or white depending on the background's luminosity.
    }
    \label{fig:note_display_modes}
\end{figure}

\para{Note display}\label{sec:note_display}
In most views, we represent notes by blocks that are positioned horizontally by start time and have a width proportional to the note's duration, to visually indicate timing and rhythm (\autoref{fig:note_display_modes}).
The first mode displays the notes as triangles in a piano roll, allowing it to represent music for any instrument, but less readable than other encodings.
A second mode displays guitar tablature, where each row stands for one string, and the third adds fret numbers for more detail.
We focus primarily on guitar tablature in this work, but new encodings resembling other instruments' common notations could extend our approach.
Depending on their size, our encodings become hard to read but still reveal coarse patterns more clearly than detailed notation.
In order to show the whole piece at once (\task{1}), without filtering or compression, we display the full score with the above encodings (\autoref{fig:teaser}g).
% Instead of common music  notation, it uses the note encodings explained in \autoref{sec:note_display} and shown in \autoref{fig:note_display_modes} for each bar.

\para{Hierarchy}
Most music pieces have a hierarchical structure in the form of sections such as verse and chorus, spanning multiple bars, each with none to a few notes, which might be grouped in harmonies (notes played at the same time).
We visualize this structure as a tree, where users can select a node to show only its children in the level below (\autoref{fig:teaser}e).
This representation supports gaining an overview (\task{1}) and allows navigating the sheet music more conveniently.
The colors are level-specific, such that they only represent similarities within, not between, the levels.
Notes have their own color map that is not based on similarity but still allows to spot repetitions or patterns such as sequences of increasing pitch (\task{2}).

\begin{figure}[htp]
  \centering
  \mbox{} \hfill
  \includegraphics[width=\linewidth]{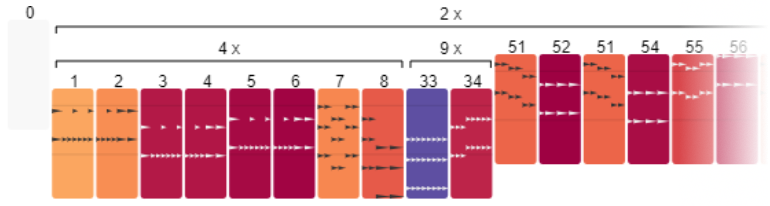}
  \hfill \mbox{}
  \caption{\label{fig:compressed}%
    Our compressed view shows repetitions as nested blocks (cut).
    Note, that bar 51 appears two times, as it equals bar 53.
   }
\end{figure}

\para{Compressed multi-level repetitions}\label{par:compressed}
Music pieces might have another hierarchical structure regarding \textit{repetitions} when a repeating segment contains repeating sub-segments (\task{1,2}).
Similar to data compression, this allows us to create a more compact representation, by only displaying a repeating segment once and annotating it with its number of repetitions.
Doing this recursively results in a tree where each leaf is a bar of the music piece, and each inner node contains the following information:
A pre-fix child, a repeated child with its repetition count, and a post-fix child, where pre- and post-fix can be empty.
The visual encoding we chose for this data structure uses our compact note encoding (\autoref{fig:note_display_modes}) for the leaves and brackets for the inner nodes (\autoref{fig:compressed}).
Numbers above the bars denote the index of their first occurrence, allowing to spot recurring ones that are farther apart.

\para{Workflow and interaction}\label{par:userWorkflow}
% user workflow
We envision musicians using our interface primarily while learning a new piece, where they first get an overview and then take a closer look at the detailed sheet music.
During navigation, reading, and playing, they could use overviews as `minimaps' that provide context for what they are currently focusing on.
As all views are linked, clicking on a bar in any view allows users to highlight or jump to a certain bar in all other views.

%%----------------------------------------------------------------------
\section{Evaluation}

We chose a pair analytics setup over a comparative user study, as most related work focuses on different tasks or data, which does not allow fair comparison.
Furthermore, instead of quantitative usability ratings and time measurements, we were more interested in qualitative feedback on \textit{how} guitarists would use our design and \textit{what limitations} they encounter.
In pair analytics~\cite{Arias2011}, designers and participants collaboratively analyze the participants' data, allowing designers to evaluate a design without needing to teach participants how to use it, saving them time and ensuring they use all features appropriately.
%
%

% participants
Our participants (P1--P4) have experience with reading guitar tablature.
P1 has played guitar for 15 years and regularly teaches it and P2 has played drums for 5 years and guitar occasionally.
P3 and P4 have played guitar for 16 and 12 years.
P1 and P3 have full color vision, P2 and P4 have a slight red-green deficiency.
All but P3 were familiar with visualization.
We met with each participant for roughly 1.5 hours.
After an introduction to our interface, we looked at guitar tablature of their choice together, encouraging them to use different features and think aloud.

% \para{Results}
% quote
\newcommand{\q}[1]{\textit{``#1''}}
% quote with participant number
\newcommand{\qp}[2]{\textit{``#1'' (P#2)}}

\begin{figure}[htp]
  \centering
  \mbox{} \hfill
  \includegraphics[width=\linewidth]{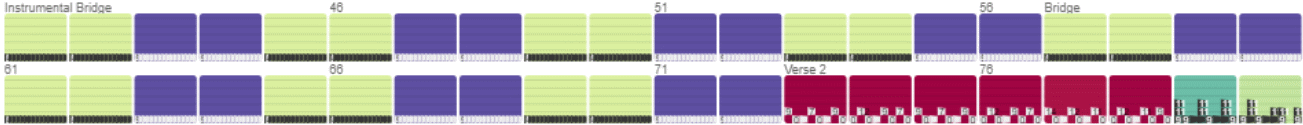}
  \hfill \mbox{}
  \caption{\label{fig:22pattern}%
   A simple alternating pattern with bars that repeat as \textit{AABB} multiple times.
   }
\end{figure}

% \paragraph{Similarity-Based Coloring}
Our participants found the coloring generally helpful:
\qp{I have played classical pieces with 8 or 12 pages ... you searched, with your teacher, made annotations with a pen ... `here it's that part again' ... if it's only black-on-white, you're blind at some point}{3}.
They were able to detect various patterns:
\qp{the color indicates a new segment}{1} (\task{1}),
\qp{always two bars one note, two bars the other note, ...}{1} (\autoref{fig:22pattern}).
One interesting example was a pattern where the same segment was repeated four times, with a different ending each time, except for the fourth that equals the second (\autoref{fig:differentendings}, \task{2, 3}).

In some cases, our current similarity metric did not work well: % or differences where barely visible:
\qp{here, I'm sure it's different, but the color is not quite different ... if you check the color carefully, I think you can see it}{1}.
We proposed coloring annotated sections of the sheet music by their similarity:
\qp{That is already very useful, because ... when I'm [teaching] and want to show segments, then I always have to mark them by hand. This is doing it for me}{1}.
While we currently colorize either by section or bar, P3 suggested alternatively coloring by sequences of bars that occur together multiple times, such as riffs or motifs.
Interestingly, our coloring allows users to quickly guess the effort needed to learn a piece:
\qp{The colors show what you already learned}{3},
\qp{For me it looks like I practice this purple bar ... and then I practice these yellow and green bars and then I can play 90 percent of the song}{1} (\task{1,2}).
P4 suggested ignoring bars with only a single note or chord when coloring, as these are less interesting.
Instead, they proposed coloring differently transposed versions of the same pattern more similarly and allowing users to manually adjust the color of a set of identical bars.

% CLUSTERING
We also compared coloring via DR versus via clustering.
When trying clustering, P3 first pondered \q{I think [coloring via clustering] is easier to understand ... as you can really see that it's different} but then concluded that
\q{it's difficult, both have pros and cons ... now if I would search [by color], it would be more difficult to spot} (\task{2,3}).
P4 suggested to additionally vary the color's brightness for different segments in the same cluster, to also reveal similarity within clusters (\task{3}).

Most patterns were visible with all color scales, although less clearly with those using fewer hues, so users with color vision deficiencies can also benefit from our design.
Even though P4 has a slight red-green deficiency, they wanted to use the default \textit{spectral} scale for most of the study.
When trying out other scales, they told us that it indeed makes a good default, as it has fewer hues than rainbow scales.
P4 further preferred \textit{viridis} over \textit{cividis}: \q{here I see better}.
When turning of colors, P3 was astonished: \q{here you see how white it is! When looking at colors for so long, you see for the first time how ugly white it is}.

The simplified and full tablature encodings still reveal characteristics:
\qp{These are power chords, right?}{2},
\qp{This is a power chord on the second fret ... and that should be A minor}{3}.
For our hierarchy view, P4 found that \q{the tab encoding doesn't help much, the simplified tab and piano roll work much better}.
Especially, since with the piano roll \qp{you can see well which bars have similar notes}{4} (\task{3}).

In our hierarchy view, clicking on different sections allows users to quickly compare them:
\qp{Main riff and verse is almost the same, it's labeled as `main riff' because it is ... without singing}{1} (\task{3}).
This also shows a drawback of sheet music, where repetition signs often apply for all instruments at once, so if one repeats and another does not, the first instrument's notes will show up redundant.
During the comparison, we found that P1 labeled the sections incorrectly, as one had a few more bars that actually belonged to the following section: \qp{we found that we labeled it wrong, that's good!}{1} (\task{3}).
Switching between different sections allows comparing them:
\qp{here's again a verse, but a little different ... back there, this bar is repeated ... this chorus is much longer}{3} (\task{2,3}).
During our study with P4, we saw that a whole bar consisted of almost only the note E, as indicated by identical color, except for a single note with different color --
\qp{I also wouldn't have noticed that [without color]}{4}.
As we only support highlighting sections and bars for now, P3 missed being able to click on single harmonies and notes to highlight them in other views.

\begin{figure}[tp]
  \centering
  \mbox{} \hfill
  \includegraphics[width=\linewidth]{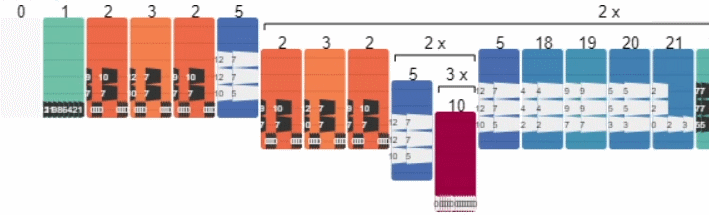}
  \hfill \mbox{}
  \caption{\label{fig:compression_unexpected}%
   Our algorithm might not create the same structure as a musician:
   The sequence $2, 3, 2, 5$ appears twice in a row, but since the following bars of the second appearance form a longer repetition, the sequence was included there instead.
   }
\end{figure}

The compressed encoding (\autoref{fig:compressed}) was P1's favorite:
\q{This is the feature I'm most excited about for showing people the song structure because this is something I just can't do with a [music notation software]} (\task{1}).
For playing a song with students where each plays a different instrument of the piece, it \q{would be great if everyone would have something like this} to have a compact summary of the part they should play.
Our participants proposed features we could add to this view, such as reducing or disabling nested repetitions:
P1 found \q{it would be great if you had the option to simplify it} and P2 suggested adding a slider for the compression level.
They also told us that this view could be extended to support annotation:
\qp{That would be great, to be able to go here and annotate some things}{1}.
Our compression sometimes leads to unexpected results as it tries to find the longest repetitions, which might not be how musicians would compress a piece:
\qp{I would expect that this is in here two times, but somehow it's here -- it makes sense, it's just two different ways of describing it}{1} (\autoref{fig:compression_unexpected}).
P3 and P4 did not consider the compressed view useful (\qp{I would not use compressed much, maybe once when first looking at a new song}{3}) and P4 proposed merging it with the compact view by optionally drawing repeated segments only once with the brackets used in this view.

% \paragraph{Compact Sheet Music}
As our compact sheet music view shows the whole piece at once, it allows users to spot global patterns, such as how often a bar appears (\task{1,2}):
\qp{The compact view shows how things repeat}{4},
\qp{If you go here and go for [the color mode] `Identical', you can see that there is a lot of this}{1}.
P4 wished for an alignment, to have repeating patterns exactly below each other: % (see \autoref{fig:differentendings} for an example without alignment):
\q{it could auto-align it for me}.
We then asked what they think about manually inserting line breaks, whereto P4 responded:
\q{this would be the most important to me -- if this should help me learn or read or play, I have to be able to customize and save it}.

When asked for general feedback, P1 told us they \q{like it a lot, because it's always hard to \emph{see} what is similar to something else ... I think it's very important that you [know] not just what is played, but also if there is a connection to other segments} (\task{1,2,3}).
P1 wished to be able to directly compare bars or sections (\task{3}): \q{That would be great if you could select two and then see the difference because I always click [back and forth]}.

When asked for use cases, our participants told us they would use our interface for learning, for example by playing along.
Both P3 and P4 further imagined using our compact view as a cheat sheet during a performance:
\qp{[For songs with chords and simpler rhythm] you could print this and give it to someone ... and they could play the song ... or you use a tablet"}{3}.
P4 hid all views but tracks and compact, thereby maximizing the latter: \q{like this, I see the whole song at once ... convenient as a memory aid. Assuming I know the song already ... I see how often I have to play everything} (\task{1,2}).
As another use case, P4 wished to be able to see multiple instruments at once to be able to compare them visually.

%%----------------------------------------------------------------------
\section{Limitations and Discussion}

% only guitar and simple metric
We mainly focus on guitar tablature, which is easier to represent compactly and often features more repeating segments than other kinds of sheet music.
However, we argue that our general method of mapping similarities to color can also help with other kinds of music.
With new, specialized similarity metrics and note encodings, our approach could support non-western kinds and even music without discrete notes, as long as a piece can be segmented.
Our example design for guitar tablature and with a simple similarity metric allowed us to stay within a reasonable scope and matched our own musical expertise.

Human color vision is limited, even more with color vision deficiencies.
Our approach can add value compared to non-colored sheet music for everyone, although accessible color scales reveal less detail.
In our study, some patterns were clearly visible while some were harder to spot -- still, they were easier to spot than without any color.
Coloring by similarity works well for pieces with a few different segments that are repeated, as fewer colors are necessary, but will not work as well for others.

As our current approach depends on dimensionality reduction and clustering, it inherits the limitation of these techniques, such as distortion and artifacts.
We chose MDS and hierarchical agglomerative clustering to preserve similarities as well as possible, but other algorithms or approaches might further reduce these limitations.

In our current interface, the participants missed being able to directly compare two selections of bars, align bars automatically or through line breaks, and assign custom colors and labels.
Our evaluation only included four participants.
While such a number is typical for pair analytics, real-world acceptance can only be evaluated through longitudinal field studies, where a larger number of users regularly use a product in their daily life.

%%----------------------------------------------------------------------
\section{Conclusion}

We designed multiple methods to ease the detection of repeating structures in sheet music.
Our evaluation provided a first qualitative indication of the effectiveness of our approach.
Therefore, we are confident that extensions to our design can turn our work into a helpful tool for musicians.

% future work
Future work includes further similarity metrics and visual encodings better suited for different tasks, sheet music characteristics, instruments, and music genres.
Adding labels and exporting them would allow musicians and teachers to save and share their results.
Showing multiple instruments of a piece at once would allow comparing them, for example, to quickly see where two guitars play similar notes.
We plan to let more musicians actively use our design during learning, playing, and teaching over months to test real-world usage and acceptance longitudinally.

%% if specified like this, the section will be committed in review mode
\section{Acknowledgments}
This work was funded by the Cyber Valley Research Fund and by the Deutsche Forschungsgemeinschaft (DFG, German Research Foundation) -- Project-ID 251654672 -- SFB TRR 161, project A08.

% For BibTeX users:
% \bibliography{ISMIRtemplate}
\bibliography{paper}

\end{document}